\documentclass[10pt,aps,pra,twocolumn,longbibliography,amsmath,amssymb]{revtex4-2}
\usepackage{graphicx}
\usepackage{dcolumn}
\usepackage{bm}
\usepackage{hyperref}
\usepackage{graphics}
\usepackage{slashed}
\usepackage{natbib}
\usepackage{url}
\usepackage[dvipsnames,svgnames,table]{xcolor}
\usepackage[T1]{fontenc}
\usepackage{diagbox}

\usepackage[normalem]{ulem}

\newcommand{\hk}{\hat{k}}
\newcommand{\vd}{v_{\mathrm{d}}}

\newcommand{\tr}{\mathrm{tr}}

\begin{document}

\title{Propagation-based classification of linear magnetoelectric response in dielectrics}%

\author{Eduardo Bittencourt$^{1,2}$}
\email{bittencourt@unifei.edu.br}
\author{Elliton O.S.R. Brand\~ao$^{1}$}
\email{ribeirobrandao@unifei.edu.br}
\author{\'Erico Goulart$^{3}$}
\email{egoulart@ufsj.edu.br}
\author{Danilo H. Spadoti$^{1}$}
\email{spadoti@unifei.edu.br}
\affiliation{$^{1}$Institute of Physics and Chemistry,  Federal University of Itajub\'a, Av. BPS, 1303, Pinheirinho, Itajub\'a/MG - Brazil}
\affiliation{$^{2}$Department of Mathematics and Applied Mathematics, University of Cape Town, Rondebosch 7700, Cape Town, South Africa}
\affiliation{$^{3}$Federal University of S\~ao Jo\~ao d'el-Rei, C.A.P. Rod.: MG 443, KM 7, Ouro Branco-MG, 36420-000, Brazil}

\date{\today}

\begin{abstract}
We study electromagnetic wave propagation in homogeneous dielectrics endowed with a linear magnetoelectric (ME) response in the geometric-optics regime. Assuming isotropic permittivity and permeability while keeping a generic $3\times 3$ ME matrix $\alpha_{ij}$, we derive the eikonal (Fresnel) eigenvalue problem for the polarization vector and obtain a compact quartic dispersion relation for the normalized phase speed $r=v/v_d$, where $v_d=(\mu\varepsilon)^{-1/2}$ is the phase speed of the underlying dielectric. We then classify the propagation effects of $\alpha_{ij}$ by decomposing it into trace, symmetric-traceless, and antisymmetric sectors. We show that (i) the pure-trace sector is propagation-silent at leading geometric-optics order; (ii) the antisymmetric sector yields a factorized quartic and produces two branches with closed-form phase speeds, including regimes where $|v|>v_d$; and (iii) the symmetric-traceless sector encodes the richest directional dependence through algebraic invariants that control the Fresnel wave surface and polarization mixing. Finally, we discuss how the predicted phase-speed shifts can be accessed by phase-sensitive transmission and resonant techniques, and we outline numerical workflows to validate the analytic dispersion and map polarization signatures in bulk and finite geometries.
\end{abstract}

\maketitle

\section{Introduction}

Linear magnetoelectricity belongs to the broader class of \emph{bianisotropic} media, where the electromagnetic excitations
$(D_{i},\, H_{i})$ depend linearly on both field strengths $(E_{i}, \,B_{i})$.
Such cross-couplings are central both to condensed-matter realizations (e.g., ME and multiferroic compounds) and to engineered metamaterials designed for polarization control and nonreciprocal functionalities. For a modern review emphasizing phenomenology and material platforms, see, e.g., \citet{Fiebig_2005}. Recent progress in optical Tellegen metamaterials further illustrates how large effective ME couplings can be achieved at visible frequencies \citep{SafaeiJazi2024NatComm}.

Historically, the linear ME effect was anticipated on symmetry grounds in antiferromagnets by Dzyaloshinskii~\cite{Dzyaloshinskii1951}, and subsequently observed in \emph{Cr$_2$O$_3$} by Astrov~\cite{Astrov1961} and shortly thereafter by Folen, Rado, and Stalder~\cite{Folen1961}. Beyond foundational interest, magnetoelectrics have been repeatedly proposed as functional materials for sensors, transducers, and electrically controlled magnetic devices; for a survey of applications and device-oriented perspectives, see Scott~\cite{Scott2012} and Fuentes-Cobas~\cite{FUENTESCOBAS2015237}.
These developments motivate a careful analysis of how ME couplings influence wave propagation, including how they modify phase speed and polarization in a way that can be probed experimentally.

From a covariant perspective, linear media can be described by a constitutive tensor $\chi^{\mu\nu\rho\sigma}$ mapping the field strength to the excitation; in \emph{premetric electrodynamics} one obtains a quartic Fresnel equation for the wave covector $k_\mu$ whose coefficients are built from $\chi^{\mu\nu\rho\sigma}$ \citep{HehlObukhov2003book,hehl2016generally,hehl2016kottler,itin2015skewon,itin2010dispersion,LammerzahlHehl2004PRD}.
A key structural result is the irreducible decomposition of the constitutive tensor into a \emph{principal} part (20 components), an \emph{axion} part (1 component), and a \emph{skewon} part (15 components). In particular, the axion contribution is known to drop out of the local Fresnel equation (it is ``propagation-silent'' in geometric optics), whereas the principal part controls birefringence and the skewon sector is tied to dissipation/nonreciprocity \citep{HehlObukhov2003book,hehl2016generally,hehl2016kottler,itin2015skewon,itin2010dispersion,LammerzahlHehl2004PRD}.

A subtle but important point in linear magnetoelectricity is the distinction between \emph{reciprocal} (lossless, energy-conserving) and \emph{nonreciprocal} (Tellegen-type) couplings.
In covariant language, energy conservation and reciprocity are naturally expressed as symmetries of $\chi^{\mu\nu\rho\sigma}$. In particular, the absence of skewon contributions is tied to a symmetric constitutive mapping and typically excludes nonreciprocal (dissipative) behavior \cite{HehlObukhov2003book,itin2015skewon,Matagne2008}. In $1+3$ language, analogous restrictions appear as symmetry relations among the $3\times 3$ blocks entering the constitutive law.

A complementary geometric viewpoint is provided by effective-geometry methods: in nonlinear electrodynamics (NLED) and related theories, the characteristics of the field equations define \emph{effective optical metrics} or, more generally, quartic wave surfaces that may factorize into two metric light cones (birefringence) or remain genuinely quartic \citep{DeLorenciEtAl2000PLB,NovelloDeLorenciEtAl2000PRD,GibbonsHerdeiro2001PRD,Perlick2000book,SchellstedePerlickLammerzahl2016}.
While our focus is linear ME response in media (rather than NLED in vacuum), the underlying mathematical structure is analogous: geometric optics reduces Maxwell's system to an algebraic eigenproblem whose characteristic equation generically defines a quartic wave surface. This makes it natural to interpret ME-induced phase-speed splitting and direction-dependent propagation in the language of wave-surface geometry.

In this broader context, it is useful to recall that NLED models also display nontrivial constraints related to causality and to electric--magnetic duality. A classic analysis of duality rotations in nonlinear electrodynamics, which underlies several later developments in Born--Infeld-type effective geometries, is given by Gibbons and Rasheed~\cite{GibbonsRasheed1995NPB}. Although our present analysis concerns linear ME media, the same conceptual warning applies: phase-speed enhancement does not by itself imply superluminal signaling, and a fully causal interpretation requires accounting for dispersion and loss (group and signal velocities), which we leave for future work.

In this work we isolate the role of magnetoelectricity by considering homogeneous isotropic permittivity and permeability, while keeping a general linear ME matrix $\alpha_{ij}$. The resulting dispersion relation can be written as a quartic polynomial for the normalized phase speed $r=v/\vd$ and depends on a small set of algebraic invariants of the ME contribution. The central technical point is that, after a $1+3$ split and in the geometric-optics limit, the canonical decomposition of $\alpha_{ij}$ into its trace, symmetric-traceless, and antisymmetric parts provides a practical \emph{propagation-based classification}: each sector has a sharply distinct impact on phase speed and polarization eigenmodes. In particular, we show that the pure-trace sector is silent at leading order (mirroring axion-like behavior in the premetric decomposition), while the antisymmetric and symmetric-traceless sectors produce measurable signatures, including regimes where the phase speed exceeds the dielectric value $\vd$. 

Related wave-propagation analyses based on Fresnel surfaces and constitutive classifications have also been explored in lower-dimensional settings, for instance, in recent studies of (2+1)-dimensional electrodynamics with general linear~\cite{BittencourtGoulartBrandao2022PRA} and nonlinear constitutive laws~\cite{Eduardo2023}. Finally, it is worth emphasizing that more elaborate propagation phenomena may arise once one goes beyond the present assumptions, for instance, by incorporating nonlinear response or higher-order ME coefficients (see, for instance, Refs.\ \cite{DeLorenci2019,De_Lorenci_2022,DeLorenci2025} and references therein).

The paper is summarized as follows. In Sec.\ \ref{sec:tools}, we introduce the linear, local constitutive relations and fix our conventions for the dielectric tensors and the reciprocal cross-coupling matrix $\alpha_{ij}$. In Sec.\ \ref{sec:fresnel} we take the geometric-optics limit and reduce Maxwell's equations to a Fresnel-matrix eigenvalue problem for the polarization amplitude. This yields a compact quartic dispersion relation for the normalized phase speed $r=v/v_d$, written solely in terms of algebraic invariants of the rescaled magnetoelectric tensor $\tilde{\alpha}_{ij}$. In Sec.\ \ref{sec:alpha_decomp} we classify the propagation effects of $\alpha_{ij}$ by decomposing it into trace, antisymmetric, and symmetric-traceless sectors, and we show how each sector affects (or does not affect) the dispersion relation. In Sec.\ \ref{sec:V_phase_pol} we analyze the polarization eigenmodes associated with each physical branch, including the appearance (in some configurations) of quasi-longitudinal electric components constrained by Gauss' law, and we provide explicit expressions for the phase-speed and polarization mixing. In Sec.\ \ref{sec:estimates} we give order-of-magnitude estimates using experimentally reported magnetoelectric coefficients, and we discuss practical measurement strategies based on phase-sensitive transmission and resonant techniques.

Throughout the paper, Latin indices $i,j,k,\ldots$ run from 1 to 3, and the Einstein convention for summation over repeated indices is used. Partial derivatives with respect to spatial and time coordinates are denoted by $\partial_i$ and $\partial_t$, respectively. The three-dimensional Levi-Civita symbol $\epsilon_{ijk}$ is a completely antisymmetric quantity defined by $\epsilon_{123} = 1$. Finally, we set the vacuum speed of light to be a unit, unless otherwise stated.


\section{Constitutive relations and Maxwell system}
\label{sec:tools}

Consider a linear, local constitutive response in a charge-free material medium. Starting from the standard thermodynamic definitions \cite{landau1984} of polarization $P_{i}$ and magnetization $M_{i}$, we write
\begin{eqnarray}
    P_{i}=P_{i}^{s}+\varepsilon_{0}\,\chi_{ij}\,E_{j}+\frac{\alpha_{ij}}{\mu_{0}}\,B_{j},
    \label{eq:Polarization}\\
    M_{i}=M_{i}^{s}+\frac{\tilde{\chi}_{ij}}{\mu_{0}}\,B_{j}-\frac{\alpha_{ji}}{\mu_{0}}\,E_{j},
    \label{eq:Magnetization}
\end{eqnarray}
where $P_{i}^{s}$ and $M_{i}^{s}$ denote spontaneous polarization and magnetization, and $E_j$ and $B_j$ correspond to the electric and magnetic field strengths, respectively. The electric $\chi_{ij}$ and magnetic $\bar\chi_{ij}$ susceptibilities give the dielectric response of the medium, and $\epsilon_0$ and $\mu_0$ are the vacuum dielectric parameters. The transpose relation between the two cross-coupling blocks (the use of $\alpha_{ij}$ and $\alpha_{ji}$ in the two sectors) corresponds to a standard \emph{reciprocal} ME coupling in many conventions for bianisotropic media \cite{landau1984,Serdyukov2001,Mackay2008}.

By defining $D_{i} = \varepsilon_{ij}\,E_{j}+P_{i}$ and $H_{i}=\mu_{ij}\,B_{j}-M_{i}$, we obtain the constitutive relations
\begin{eqnarray}
    D_{i}=P_{i}^{s}+\varepsilon_{ij}\,E_{j}+\frac{\alpha_{ij}}{\mu_{0}}\,B_{j},
    \label{eq:Dvector}\\
    H_{i}=M_{i}^{s}+\mu_{ij}\,B_{j}-\frac{\alpha_{ji}}{\mu_{0}}\,E_{j},
    \label{eq:Hvector}
\end{eqnarray}
where $\varepsilon_{ij}=\varepsilon_{0}\,(\delta_{ij}+\chi_{ij})$ is the electric permittivity tensor and $\mu^{-1}_{ij}=\frac{1}{\mu_{0}}\,(\delta_{ij}-\tilde{\chi}_{ij})$ is the inverse permeability tensor. The vectors $D_i$ and $H_j$ are the field excitations, encoding the medium response to the external fields $E_i$ and $B_j$.

In this medium, Maxwell's equations read
\begin{eqnarray}
    \partial_i\,B_{i}=0,
    \label{eq:GaussB}\\
    \epsilon_{ijk}\,\partial_{j}\,E_{k} = -\partial_{t}\,B_{i},
    \label{eq:Faraday}\\
    \partial_i\,D_{i}=0,
    \label{eq:GaussD}\\
    \epsilon_{ijk}\,\partial_{j}\,H_{k} = \partial_{t}\,D_{i}.
    \label{eq:AmpereMaxwell}
\end{eqnarray}
Combining Eq.\ \eqref{eq:AmpereMaxwell} with Eqs.\ \eqref{eq:Dvector}--\eqref{eq:Hvector}, and using Eq.\ \eqref{eq:Faraday} to eliminate $B_{i}$ when appropriate, one can derive a second-order wave equation for $E_{i}$. In full generality (allowing inhomogeneous constitutive tensors) one finds the schematic structure
\begin{eqnarray}
\partial_{t}^{2}\big(\varepsilon_{ij}E_{j}\big)
+\epsilon_{ijk}\epsilon_{lmn}\,\partial_{j}\partial_{m}\big(\mu^{-1}_{kl}E_{n}\big)
=\nonumber\\[1ex]
-\frac{1}{\mu_{0}}\partial_{t}\left[\epsilon_{ijk}\,\partial_{j}(\alpha_{lk}E_{l})
-\epsilon_{jkl}\,\partial_{k}(\alpha_{ij}E_{l})\right],
\label{eq:waveeqE}
\end{eqnarray}
which simplifies considerably for homogeneous media, where $\varepsilon_{ij},\mu_{ij},\alpha_{ij}$ are constants.

A general second-order equation solely for $B_{i}$ is not always convenient unless additional constraints are imposed on the constitutive tensors. However, in geometric optics, the Faraday law (\ref{eq:Faraday}) enforces the familiar mutual orthogonality relations between $E_{i}$, $B_{i}$, and the wave covector, and the propagation problem reduces to an algebraic eigenvalue system for the polarization.

\section{Geometric-optics reduction and Fresnel matrix}
\label{sec:fresnel}

In order to singularize the effects of linear ME coefficients on light propagation in the geometric optics regime, we choose the dielectric parameters to be homogeneous and isotropic, namely, $\varepsilon_{ij}=\varepsilon\,\delta_{ij}$ and $\mu^{-1}_{ij}=\mu^{-1}\,\delta_{ij}$. 

Under the standard eikonal \textit{Ansatz}
\begin{equation}
E_j = \Re\Big\{ e_j\,\exp\big[i\,\Sigma(x_k,t)/\epsilon\big]\Big\},\qquad
\epsilon\to 0^+,
\end{equation}
with slowly varying complex amplitudes $e_j$ and rapidly varying phase $\theta$, the leading order term in $\epsilon$ for the wave equation \eqref{eq:waveeqE} yields an algebraic system
\begin{equation}
\mu\varepsilon\,\omega^2\, e_i
-k^2 e_i+k_i k_j e_j
+ \mu_r\,\omega k_j\,\epsilon_{jk(i}\,\alpha_{l)k}\,e_l
=0,
\label{eq:eikonal_basic}
\end{equation}
where we define the wave covector $k_i = \partial_i\Sigma$, frequency $\omega = -\partial_t\Sigma$, and $\mu_r=\mu/\mu_0$ is the relative permeability. The parentheses denote symmetrization, hence
$\epsilon_{jk(i}\alpha_{l)k}=\epsilon_{jki}\alpha_{lk}+\epsilon_{jkl}\alpha_{ik}$.

Dividing Eq.\ \eqref{eq:eikonal_basic} by $k^2\equiv k_ik_i$ and using the definition of phase speed $v=\omega/k$, we obtain the Fresnel-matrix eigen-problem
\begin{equation}
Z_{il} e_l=0,
\label{eq:Zdef}
\end{equation}
with
\begin{equation}
Z_{il}=\mu\varepsilon\,v^2\,\delta_{il}+\mu_r\,v\,\hat{k}_j\,\epsilon_{jk(i}\,\alpha_{l)k}-I_{il},
\end{equation}
where $I_{il}=\delta_{il}-\hat{k}_i\hat{k}_l$ is the Euclidean projector onto the plane orthogonal to $\hat{k}_{i}=k_i/k$. Thus, the dispersion relation is the condition for non-trivial polarization solutions
\begin{equation}
\det(Z_{il})=0.
\label{eq:detZ0}
\end{equation}

For later convenience, we define the \textit{dielectric phase speed} $\vd\equiv(\mu\varepsilon)^{-1/2}$ and the ratio
$r\equiv v/\vd$. With these elements, we introduce the rescaled matrix 
\begin{equation}
\label{eq:alphatilde_def}
\tilde{\alpha}_{ij}\equiv \mu_r\,\vd\,\hat{k}_l\,\epsilon_{lk(i}\,\alpha_{j)k}
\end{equation}
and the directional scalar $\tilde P \equiv \hat{k}_i\,\tilde{\alpha}_{ik}\,\tilde{\alpha}_{kj}\,\hat{k}_j$. Then, Eq. \eqref{eq:detZ0} can be written as a quartic polynomial in $r$,
\begin{equation}
\label{eq:char_poly}
r^4+a_3 \, r^3+a_2r^2
+a_1 r+a_0=0,
\end{equation}
with
\begin{equation}
\begin{array}{lcl}
a_3&=&\tilde\alpha\equiv\tr(\tilde\alpha^{i}{}_{j}), \\[1ex]
a_2&=&\frac{1}{2}\Big(\tilde\alpha^2 - \tilde\alpha_{ij}\tilde\alpha_{ji}-4\Big),\\[1ex]
a_1&=&\det(\tilde\alpha_{ij})-\tilde\alpha,\\[1ex]
a_0&=&\tilde P+1.
\end{array}
\end{equation}
Equation \eqref{eq:char_poly} encodes up to two physical propagating branches (two real $v>0$) for given $(\hk_i,\alpha_{ij})$, and provides a natural starting point for classification via algebraic invariants. It is important to emphasize that, written in this way, the characteristic polynomial \eqref{eq:char_poly} accounts for the ME effects on the light propagation without the necessity of quantifying the dielectric counterpart.

In premetric electrodynamics, the Fresnel equation defines a quartic wave surface (Tamm-Rubilar surface) whose factorization properties determine whether the medium is non-birefringent (double light cone) or birefringent (two distinct cones) \citep{HehlObukhov2003book,Itin2004CFJ,LammerzahlHehl2004PRD}.
Effective-geometry methods in NLED similarly show that characteristics can be described by quartic structures and that special algebraic conditions lead to factorization into two optical metrics \citep{DeLorenciEtAl2000PLB,NovelloDeLorenciEtAl2000PRD,GibbonsHerdeiro2001PRD,SchellstedePerlickLammerzahl2016}.
Our $1+3$ formulation provides a concrete realization of these ideas in the ME sector: the matrix $\tilde{\alpha}_{ij}$ is the only ME object entering the geometric-optics Fresnel problem, and it depends on $\alpha_{ij}$ only through specific contractions with $\hk_i$ and $\epsilon_{ijk}$, cf. Eq.\ \eqref{eq:alphatilde_def}. Therefore, this makes the canonical $3\times 3$ decomposition of $\alpha_{ij}$ into trace, symmetric-traceless, and antisymmetric parts a natural \emph{propagation classifier}.

\section{Decomposing the ME tensor}
\label{sec:alpha_decomp}

From the algebraic point of view, the ME tensor $\alpha_{ij}$ decomposes uniquely as
\begin{equation}
\alpha_{ij}=\frac{1}{3}\alpha\,\delta_{ij}
+\alpha^{(\mathrm{ST})}_{ij}
+\alpha^{(\mathrm{A})}_{ij},
\label{eq:alpha_decomp}
\end{equation}
with
\begin{equation}
\alpha:=\alpha^k{}_k,
\qquad
\alpha^{(\mathrm{ST})}_{ii}=0,
\qquad
\alpha^{(\mathrm{A})}_{ij}=-\alpha^{(\mathrm{A})}_{ji}.
\end{equation}
We shall see that it is natural to interpret this splitting of $\alpha_{ij}$ not merely as a mathematical convenience, but as a practical proxy for classifying the propagation-relevant components of the reciprocal ME response, by showing how each sector affects (or does not affect) the dispersion relation given by Eq.\ \eqref{eq:char_poly}.

\subsection{Pure trace}
If $\alpha_{ij}=\tfrac{\alpha}{3}\delta_{ij}$, then $\tilde{\alpha}_{ij}=0$ because $\epsilon_{jk(i}\delta_{l)k}=0$ by the antisymmetry of $\epsilon_{jkl}$. Hence, $\tilde\alpha_{ij}=0$ and Eq.\ \eqref{eq:char_poly} reduces to
\begin{equation}
(r^2-1)^2=0
\qquad\Rightarrow\qquad
v=\vd.
\label{eq:pure_trace_result}
\end{equation}
Thus, the pure-trace ME sector is propagation-silent at leading geometric-optics order, in close analogy with the axion sector in the premetric irreducible decomposition \citep{HehlObukhov2003book,Itin2004CFJ}.

\subsection{Pure antisymmetric sector}
\label{sec:antisym}
If $\alpha_{ij}$ is purely antisymmetric, it can be parameterized by a pseudovector $\alpha_{k}$,
\begin{equation}
\label{eq:param_alpha}
\alpha_{ij}=\epsilon_{ijk}\,\alpha_k.
\end{equation}
By defining the directional scalars $\beta\equiv \alpha_i\hk_i$ and $\alpha^2\equiv \alpha_i \alpha_i$, a direct computation gives
\begin{equation}
\tilde\alpha_{ij}=\mu_r\,v_d (\alpha_i\hk_j+\alpha_j\hk_i-2\beta\,\delta_{ij}).
\label{eq:param_alpha_tilde}
\end{equation}
The invariants entering in Eq.\ \eqref{eq:char_poly} can then be expressed in terms of rescaled $\tilde\alpha=\mu_r\,v_d\,\alpha$ and $\tilde\beta=\mu_r\,v_d\,\beta$, leading to a \emph{factorized} quartic:
\begin{eqnarray}
\label{eq:char_poly_skew}
\Big[(r-\tilde\beta)^2-\tilde\beta^2-1\Big]\Big[(r-\tilde\beta)^2-\tilde\alpha^2-1\Big]=0,
\end{eqnarray}
so that the phase-speed branches are
\begin{equation}
v=\vd\Big(\tilde\beta\pm\sqrt{1+\tilde\beta^2}\Big)
\quad\text{or}\quad
v=\vd\Big(\tilde\beta\pm\sqrt{1+\tilde\alpha^2}\Big).
\label{eq:antisym_speeds}
\end{equation}
Two instructive special cases are: (i) $\beta=0$ (propagation orthogonal to $\alpha_{k}$), and (ii) $\alpha^2=\beta^2$ (propagation parallel to $\alpha_{k}$). Equation \eqref{eq:antisym_speeds} also makes explicit that, for suitable signs and parameters, one can obtain $|v|>\vd$ even though $\varepsilon$ and $\mu$ are unchanged. As usual in media, a phase-speed enhancement does not by itself imply superluminal signal propagation; group velocity, dispersion, and causality constraints must be addressed once frequency dependence and losses are included (cf. the general causality analyses in effective-geometry approaches \citep{GibbonsHerdeiro2001PRD,SchellstedePerlickLammerzahl2016}).

\subsection{Pure symmetric-traceless sector}
\label{sec:sytr}
Let $\alpha_{ij}=\alpha_{ji}$ with $\alpha^i{}_i=0$. By the spectral theorem, there exists an orthonormal basis in which
\begin{equation}
\label{alpha_sym_spectral}
\alpha_{ij}=\mathrm{diag}(\lambda_1,\lambda_2,\lambda_3),\qquad \lambda_3=-(\lambda_1+\lambda_2).
\end{equation}
Because the dielectric sector is isotropic, this basis choice does not alter the expressions for $\varepsilon_{ij}$ and $\mu_{ij}=$.
On this basis, the invariants in Eq.\ \eqref{eq:char_poly} reduce to explicit functions of $(\lambda_1,\lambda_2)$ and the direction cosines $\hk_i$. For instance,
\begin{equation}
\begin{array}{l}
\tilde\alpha_{ij}\tilde\alpha_{ij}= 2\Big\{3\tilde\lambda_1^2\hk_2^2+3\tilde\lambda_2^2\hk_1^2 +2\tilde\lambda_1\tilde\lambda_2\big[3(\hk_1^2+\hk_2^2)-1\big]\Big\},\nonumber\\[1ex]
\det(\tilde\alpha^i{}_j)=-2\,\hk_1\hk_2\hk_3(\tilde\lambda_1-\tilde\lambda_2)(\tilde\lambda_1+2\tilde\lambda_2)(\tilde\lambda_2+2\tilde\lambda_1),\nonumber\\[1ex]
\tilde P=\tilde\lambda_1^2\hk_1^2+\tilde\lambda_2^2\hk_2^2+\tilde\lambda_3^2\hk_3^2-\big(\tilde\lambda_1\hk_1^2+\tilde\lambda_2\hk_2^2+\tilde\lambda_3\hk_3^2\big)^2,\nonumber
\end{array}
\end{equation}
where $\tilde\lambda_a=\mu_r\,\vd\,\lambda_a$, with $a=1,2,3$. This sector is therefore the most sensitive to angular scans: varying the components of $\hk_i$ changes the algebraic invariants controlling \eqref{eq:char_poly}, producing a characteristic directional phase-speed ``fingerprint'' and typically a polarization splitting akin to biaxial birefringence.

\section{Phase speed and polarization analysis}
\label{sec:V_phase_pol}

In the geometric-optics regime, once a root $v$ (or $r=v/\vd$) of the dispersion relation is selected, the polarization vector $e_j$
is obtained from the null space of the Fresnel matrix,
\begin{equation}
Z_{ij}(v,\hk_l)\,e_j=0.
\end{equation}
The associated magnetic amplitude follows from Faraday's law,
\begin{equation}
b_{l}=\frac{1}{v}\,\epsilon_{lmn}\,\hat{k}_{m}\,e_{n},
\label{eq:b_from_e}
\end{equation}
so that $b_{i}\hk_{i}=0$, identically. In a ME medium, Gauss' law \eqref{eq:GaussD} constrains the longitudinal component of $e_{i}$
through the constitutive relation $D_{i}=\varepsilon_{ij}\,E_{j}+\mu_{0}^{-1}\alpha_{ij}B_{j}$, so a nonzero longitudinal component $e_{i}\hk_{i}$ may appear in some branches, vanishing in the purely dielectric case. Notice that the eigenproblem always contains a spurious longitudinal solution at $r=0$ associated with non-propagating modes, which can be excluded in favor of the physical branches $v\neq0$.

\subsection{Pure trace (propagation-silent)}
\label{subsec:pol_trace}

For $\alpha_{ij}=\tfrac{\alpha}{3}\delta_{ij}$ one has $\tilde\alpha_{ij}=0$, hence
\begin{equation}
Z_{ij}=r^2\delta_{ij}-I_{ij}.
\end{equation}
The dispersion condition $\det Z=0$ gives $(r^2-1)^2=0$, i.e. $r=\pm 1\,\Rightarrow\,v=\pm\vd$. The polarization equation becomes
\begin{equation}
(r^2\delta_{ij}-I_{ij})e_j=0.
\end{equation}
For the physical roots $r^2=1$, this implies
\begin{equation}
I_{ij}e_j=e_i
\qquad\Leftrightarrow\qquad
\hk_{i}e_{i}=0,
\end{equation}
so the electric field is purely transverse and \emph{degenerate}: any pair of orthonormal vectors spanning the plane orthogonal to $\hk_{i}$
provides a valid basis of polarization states. Consequently, in the pure-trace ME sector there is no phase-speed splitting and no preferred
polarization axis at leading geometric-optics order.

\subsection{Purely antisymmetric}
\label{subsec:pol_antisym}

In this case, the ME tensor can be parameterized according to Eqs.\ \eqref{eq:param_alpha}-\eqref{eq:param_alpha_tilde} and the rescaled (dimensionless) invariants $\tilde{\beta}$ and $\tilde{\alpha}$. Thus, we decompose $\alpha_i$ into components parallel and orthogonal to $\hat{k}_i$, 
\begin{equation}
\alpha_i=\beta\,\hat{k}_i+\alpha^{\perp}_i,
\quad
\alpha^\perp_i\hat{k}_i=0,
\quad
(\alpha^\perp)^2:=\alpha^\perp_i\alpha^\perp_i=\alpha^2-\beta^2.
\end{equation}
We also define the corresponding rescaled transverse magnitude
\begin{equation}
\tilde{\alpha}_\perp:=\mu_r\,\vd\,\sqrt{\alpha^2-\beta^2}
=\sqrt{\tilde{\alpha}^2-\tilde{\beta}^2}.
\end{equation}
When $\alpha^\perp_i\neq0$ (equivalently $\tilde{\alpha}_\perp\neq0$), it is convenient to introduce the orthonormal triad $\{\hat{u}_i,\hat{v}_i,\hat{k}_i\}$ adapted to the geometry of the problem given by
\begin{equation}
\hat{u}_i:=\frac{\alpha^\perp_i}{\sqrt{\alpha^2-\beta^2}},
\qquad
\hat{v}_i:=\epsilon_{ijm}\,\hat{k}_j\,\hat{u}_m.
\label{eq:triad_u_v_k}
\end{equation}
We then expand the polarization eigenvector as
\begin{equation}
e_i = E_u\,\hat{u}_i + E_v\,\hat{v}_i + E_\parallel\,\hat{k}_i.
\label{eq:e_expand_uvk}
\end{equation}
In this basis, the Fresnel eigenproblem $Z_{ij}e_j=0$ decouples into two independent branches.

\paragraph{Branch (I): purely transverse mode orthogonal to $\alpha^\perp_i$.}
The component along $\hat{v}_i$ is an eigen-direction of $Z_{ij}$ and satisfies
\begin{equation}
\big(r^2-1-2\tilde{\beta}\,r\big)\,E_v=0,
\qquad\Longrightarrow\qquad
r=\tilde{\beta}\pm\sqrt{1+\tilde{\beta}^2}.
\label{eq:r_branchI_tildebeta}
\end{equation}
The corresponding eigenpolarization is
\begin{equation}
e^{(I)}_i \ \propto\ \hat{v}_i,
\qquad
\hat{k}_i e^{(I)}_i=0,
\qquad
\hat{u}_i e^{(I)}_i=0,
\label{eq:pol_branchI}
\end{equation}
hence Branch (I) is strictly transverse and linearly polarized, with direction fixed by the plane spanned by $\hat{k}_i$ and $\alpha_i$.

\paragraph{Branch (II): hybrid mode.} The remaining $(\hat{u}_i,\hat{k}_j)$ components form a coupled $2\times2$ system whose compatibility condition gives
\begin{equation}
r^2-2\tilde{\beta}\,r-(1+\tilde{\alpha}^2-\tilde{\beta}^2)=0
\qquad\Longrightarrow\qquad
r=\tilde{\beta}\pm\sqrt{1+\tilde{\alpha}^2}.
\label{eq:r_branchII_tildealpha}
\end{equation}
For any such root with $r\neq 0$, one may choose the eigenvector in the compact form
\begin{equation}
e^{(II)}_i \ \propto\ \hat{u}_i-\frac{\tilde{\alpha}_\perp}{r}\,\hat{k}_i,
\label{eq:pol_branchII}
\end{equation}
so that the longitudinal-to-transverse ratio becomes
\begin{equation}
\frac{\hat{k}_i e^{(II)}_i}{\hat{u}_i e^{(II)}_i}
=-\frac{\sqrt{\tilde{\alpha}^2-\tilde{\beta}^2}}{r}.
\label{eq:long_ratio_antisym_tilde}
\end{equation}
Thus, Branch (II) is a hybrid (quasi-longitudinal) polarization, in general, where a longitudinal component is present unless
$\alpha^\perp_i=0$.

Finally, two special configurations warrant attention. The first one corresponds to $\alpha^\perp_i=0$ (equivalently $\alpha_i\parallel\hat{k}_i$, hence $\tilde{\alpha}_\perp=0$ and $\tilde{\alpha}^2=\tilde{\beta}^2$), where the two branches coincide, and the medium is effectively non-birefringent in geometric optics. Then, any transverse polarization is allowed, namely, degeneracy restored. The second case is if $\beta=0$ (equivalently $\alpha_i\perp\hat{k}_i$, hence $\tilde{\beta}=0$ and $\tilde{\alpha}_\perp=\tilde{\alpha}$), then Branch (I) reduces to $r=\pm 1$ (dielectric speed) with $e_i\parallel\hat{v}_i$, while Branch (II) yields
$r=\pm\sqrt{1+\tilde{\alpha}^2}$ with hybrid polarization \eqref{eq:pol_branchII}. In this regime one obtains the explicit phase-speed enhancement
\begin{equation}
\label{eq:anti_enh_beta0}
|v|=\vd\,\sqrt{1+\tilde{\alpha}^2}>\vd
\end{equation}
on Branch (II), at fixed $(\varepsilon,\mu)$.

\subsection{Symmetric-traceless}
\label{subsec:pol_symtr_hk3zero}

We now specialize to the traceless symmetric sector and analyze propagation in the plane perpendicular to the third eigen-direction, i.e. $\hk_3=0$. With the eigenbasis of $\alpha_{ij}$ that gives Eq.\ \eqref{alpha_sym_spectral}, we parameterize the propagation direction by an angle $\phi$ in the $(\hat k_1,\hat k_2)$-plane, as follows
\begin{equation}
\hk_i=(\cos\phi,\sin\phi,0).
\end{equation}
Now, with the dimensionless eigenvalues $\tilde\lambda_a$ we define the convenient combinations
\begin{eqnarray}
&&\Delta_1:=\frac{\tilde\lambda_3-\tilde\lambda_1}{2},
\qquad
\Delta_2:=\frac{\tilde\lambda_3-\tilde\lambda_2}{2},\nonumber\\[1ex]
&&\Delta_{12}:=\Delta_1-\Delta_2=\frac{\tilde\lambda_2-\tilde\lambda_1}{2}.
\label{eq:Deltas_def}
\end{eqnarray}

\subsubsection{Phase-speed equation for $\hk_3=0$}

For $\hk_3=0$, the Fresnel determinant becomes an even polynomial, and the physical dispersion reduces to a quadratic equation for $x\equiv r^2$. A compact derivation follows by working in the orthonormal basis $\{\hat k_i,\hat t_j,\hat e^{(3)}_k\}$, where
\begin{equation}
\hat{t}_i:=(-\sin\phi,\cos\phi,0),\quad\mbox{and}\quad
\hat{e}^{(3)}_i:=(0,0,1),
\end{equation}
being $\hat t_i$ in-plane transverse and $\hat{e}^{(3)}_i$ out-of-plane transverse directions. By writing the polarization as
\begin{equation}
e_j=E_t\,\hat{t}_j+E_3\,\hat{e}^{(3)}_j+E_\parallel\,\hk_j,
\label{eq:pol_decomp_symtr}
\end{equation}
we find that the Fresnel system $Z_{ij}e_j=0$ is equivalent to the scalar compatibility condition
\begin{equation}
(x-1)^2-x\,G(\phi)^2-(x-1)\,Q(\phi)=0,
\label{eq:disp_symtr_plane_compact}
\end{equation}
where
\begin{equation}
G(\phi)=\Delta_2\,\cos^2\phi+\Delta_1\,\sin^2\phi,
\qquad
Q(\phi)=\frac{\Delta_{12}^2}{4}\,\sin^2(2\phi).
\label{eq:GQ_def}
\end{equation}
Equivalently, Eq.\ \eqref{eq:disp_symtr_plane_compact} can be written as a quadratic in $x$, as follows
\begin{equation}
x^2-\big(2+G^2+Q\big)\,x+(1+Q)=0,
\label{eq:disp_symtr_plane_quadratic}
\end{equation}
with solutions
\begin{equation}
r^2_{\pm}(\phi)=
\frac{2+G^2+Q\ \pm\ \sqrt{\big(2+G^2+Q\big)^2-4\big(1+Q\big)}}{2}.
\label{eq:r2_pm_symtr_plane}
\end{equation}
The corresponding phase speeds are $v_\pm(\phi)=\vd\,r_\pm(\phi)$, choosing the physical sign $v>0$.

Along principal directions in the plane, we find
\begin{align}
\phi=0:\quad &G=\Delta_2,\ Q=0\ \Rightarrow\ x^2-(2+\Delta_2^2)x+1=0,\\
\phi=\frac{\pi}{2}:\quad &G=\Delta_1,\ Q=0\ \Rightarrow\ x^2-(2+\Delta_1^2)x+1=0.
\end{align}
Thus, one obtains two branches with $r_+>1>r_-$ (one enhanced and one reduced phase speed) for generic $\Delta_{1,2}\neq0$.

In the uniaxial limit $\lambda_1=\lambda_2$, we get $\Delta_{12}=0$ and, consequently, $Q(\phi)=0$ for all $\phi$. Therefore, the dispersion relation reduces to
\begin{equation}
\label{eq:uni_lim_trless_dis_rel}
x^2-\big(2+G^2\big)x+1=0,
\end{equation}
with $G\equiv \Delta_1\equiv\Delta_2$. So, the splitting becomes independent of the in-plane angle $\phi$, leading to an axial symmetry around $\hat{\bm{e}}_3$.

Recall that ``dielectric-speed'' directions ($r^2=1$) are allowed only if $G(\phi)=0$. From Eq.\ \eqref{eq:GQ_def}, this requires $\Delta_1$ and $\Delta_2$ to have opposite signs, and a specific angle $\phi$ such that $\Delta_2\cos^2\phi+\Delta_1\sin^2\phi=0$. Along such directions, one branch propagates at $v=\vd$ with a purely transverse in-plane polarization, as we shall see below.

\subsubsection{Polarizations for each in-plane branch}

For any physical root $r(\phi)$ of Eq.\ \eqref{eq:r2_pm_symtr_plane}, the Fresnel equations determine the ratios among $(E_t,E_3,E_\parallel)$ in Eq.\ \eqref{eq:pol_decomp_symtr}. A convenient parametrization is obtained by choosing $E_3\neq0$ (generic case) and expressing the remaining components in terms of it, as follows
\begin{equation}
E_t=\frac{r\,G(\phi)}{r^2-1}\,E_3,
\qquad
E_\parallel=-\frac{\cos\phi\,\sin\phi\,\Delta_{12}}{r}\,E_3.
\label{eq:Et_Epar_in_terms_E3}
\end{equation}
Therefore, up to an overall normalization, the eigenpolarization associated with a given branch $r=r_\pm(\phi)$ can be written as
\begin{equation}
e^{\pm}_{j}(\phi)\ \propto\
\frac{r_\pm\,G}{r_\pm^2-1}\,\hat{t}_j \;+\;\hat{e}^{(3)}_j\;-\;\frac{\cos\phi\,\sin\phi\,\Delta_{12}}{r_\pm}\,\hk_j.
\label{eq:pol_symtr_plane_final}
\end{equation}

Note that a (real) mixing angle $\psi_\pm$ in terms of the transverse part of the polarization can be defined as
\begin{equation}
\tan\psi_\pm=\frac{E_t}{E_3}=\frac{r_\pm\,G}{r_\pm^2-1}.
\end{equation}
Thus, even though both $\hat{t}_j$ and $\hat{e}^{(3)}_j$ are transverse to $\hk_i$, the ME coupling fixes their relative weight, selecting preferred polarization directions.

From Eq.\ \eqref{eq:Et_Epar_in_terms_E3}, we see that the longitudinal component is controlled by $\Delta_{12}$ and by the in-plane angle such that
\begin{equation}
\frac{E_\parallel}{E_3}
=-\frac{\cos\phi\,\sin\phi\,\Delta_{12}}{r_\pm(\phi)}.
\end{equation}
Hence $E_\parallel=0$ in any of the following situations: (i) propagation along the principal axes in the plane ($\phi=0$ or $\phi=\pi/2$), for which $\cos\phi\,\sin\phi=0$; (ii) the uniaxial limit $\lambda_1=\lambda_2$ ($\Delta_{12}=0$); or (iii) in the formal limit $|r|\to\infty$. In such cases, the eigenpolarizations are \emph{strictly transverse} and lie entirely in the $\{\hat{t}_i,\hat{e}^{(3)}_j\}$ plane.

Finally, if $G(\phi)=0$, one may have a solution with $r^2=1$. In that case, the first relation in Eq.\ \eqref{eq:Et_Epar_in_terms_E3} does not apply.
Instead, the Fresnel equations enforce $E_3=0$ (generic) and select a purely transverse in-plane polarization, that is
\begin{equation}
r^2=1,\ G(\phi)=0
\quad\Rightarrow\quad
e_j\ \propto\ \hat{t}_j,
\qquad
e_j\hk_j=0.
\end{equation}
This is a genuine geometric-optics analogue of an ``optical-axis'' direction within the plane, where one branch propagates at the dielectric speed.

\section{Estimates}
\label{sec:estimates}
This section provides order-of-magnitude estimates for the phase-speed modifications predicted in the previous section using experimentally reported linear ME coefficients of real materials.
Our goal is to show that the effects can be made observable with standard phase-sensitive metrology (from microwave to THz), and to provide concrete benchmark numbers for one representative material
in each nontrivial propagation-relevant sector (antisymmetric and symmetric-traceless).

Experimental papers often report the linear ME effect in the polarization form $P_i = \alpha^{(H)}_{ij} H_j$ (or, equivalently, the magnetization form $M_i=\alpha^{(E)}_{ij}E_j$),
with $\alpha^{(H)}_{ij}$ in SI units of $\mathrm{s/m}$ (frequently quoted in $\mathrm{ps/m}$).
From our constitutive relations, given by Eqs.\ \eqref{eq:Polarization}-\eqref{eq:Magnetization}, we obtain
\begin{equation}
P_i=\frac{1}{\mu_0}\,\alpha_{ij} B_j,
\qquad B_j=\mu_0\mu_r H_j,
\end{equation}
so that
\begin{equation}
\alpha^{(H)}_{ij}=\mu_r\,\alpha_{ij}.
\label{eq:alpha_map_H_to_B}
\end{equation}
For the vast majority of insulating MEs, one has $\mu_r\simeq 1$ at microwave frequencies \cite{Balanis2012}, hence $\alpha^{(H)}_{ij}\approx \alpha_{ij}$ to excellent accuracy. Nevertheless, Eq.~\eqref{eq:alpha_map_H_to_B} should be used if $\mu_r$ significantly differs from unity (for instance, in
engineered composites or metamaterials).

The geometric-optics dispersion depends on the \emph{dimensionless} combinations
$\tilde{\alpha}\sim \mu_r v_d \alpha$, where $v_d=(\mu\varepsilon)^{-1/2}=c/\sqrt{\mu_r\varepsilon_r}$, and the speed of light $c$ is restored for dimensional reasons. Thus, for any ME coefficient with SI units $\mathrm{s/m}$,
\begin{equation}
\mu_r v_d\,\alpha
=\frac{c\,\alpha}{\sqrt{\varepsilon_r/\mu_r}}
\simeq \frac{c\,\alpha}{\sqrt{\varepsilon_r}}
\qquad (\mu_r\simeq 1),
\label{eq:dimensionless_strength}
\end{equation}
which provides the natural figure of merit controlling the phase-speed shift.

\subsection{Measurable effects: phase delay and cavity shifts}

In a homogeneous slab of thickness $L$, a given branch with $v=r\,v_d$ produces a time-of-flight shift
relative to the underlying dielectric speed $v_d$,
\begin{equation}
\Delta t(L)=\frac{L}{v}-\frac{L}{v_d}
=\frac{L}{v_d}\left(\frac{1}{r}-1\right).
\label{eq:tof_shift}
\end{equation}
At angular frequency $\omega$, this corresponds to a directly measurable transmission phase shift
\begin{equation}
\Delta\Phi(L)=\omega\,\Delta t(L)
=\frac{\omega L}{v_d}\left(\frac{1}{r}-1\right).
\label{eq:phase_shift}
\end{equation}
Equivalently, the effective refractive index of a branch is $n=c/v=\sqrt{\varepsilon_r\mu_r}/r$,
so that small shifts obey $\Delta n/n\simeq -\Delta r/r$.

A particularly sensitive implementation is a resonant cavity \cite{Waldron1960} or a guided-wave resonator \cite{Krupka2001} filled with the ME medium. To leading order, the resonant frequency shift scales as $\Delta f/f \sim -\Delta n/n$ (up to standard filling-factor corrections), allowing ppm--ppb
resolution with high-quality factor $Q$ structures. This is especially relevant for weak single-phase ME crystals.

\subsection{Antisymmetric representative TbPO$_4$}

Balanced ME annealing of TbPO$_4$ yields a tensor with a single dominant off-diagonal component (quoted in the literature as $\alpha_{xy}$ or $\alpha_{yx}$), with reported peak values as large as $\alpha \sim 730~\mathrm{ps/m}$ at low temperature ($T\sim 1$--$2~\mathrm{K}$) \cite{Rado1984TbPO4,Rivera2009,Shen2019DyCrO4}.
In the language of Sec.\ \ref{sec:antisym}, this situation is naturally modeled by a purely antisymmetric sector $\alpha_{ij}=\epsilon_{ijk}\alpha_k$ with $\alpha_k$ along the tetragonal axis. Choosing propagation orthogonal to $\alpha_k$ implies $\beta=\alpha_i\hat{k}_i=0$, so that Eq.\ \eqref{eq:anti_enh_beta0} applies and the enhanced branch obeys
\begin{equation}
|v|=v_d\sqrt{1+\tilde{\alpha}^2},
\qquad
\tilde{\alpha}\simeq \frac{c\,\alpha}{\sqrt{\varepsilon_r}}.
\label{eq:TbPO4_enhancement}
\end{equation}
Taking $\mu_r\simeq 1$ and a conservative oxide-scale $\varepsilon_r\simeq 10$ for an order-of-magnitude
estimate, one finds
\begin{equation}
\tilde{\alpha}\ \approx\ \frac{(3.0\times 10^8~\mathrm{m/s})(7.3\times 10^{-10}~\mathrm{s/m})}{\sqrt{10}}
\ \approx\ 6.9\times 10^{-2},
\end{equation}
with
\begin{equation}
\frac{|v|}{v_d}-1\ \approx\ 2.4\times 10^{-3}.
\end{equation}
Hence, the antisymmetric ME coupling in TbPO$_4$ can produce a \emph{quarter-percent} phase-speed
enhancement on the hybrid branch of Sec.\ \ref{subsec:pol_antisym}.

As a concrete microwave figure, at $f=10~\mathrm{GHz}$, one has
$v_d\simeq c/\sqrt{10}\approx 9.5\times 10^7~\mathrm{m/s}$, and Eq.~\eqref{eq:phase_shift} gives
\begin{equation}
\Delta\Phi(L)\ \approx\ \frac{2\pi f\,L}{v_d}\left(\frac{1}{1.0024}-1\right)
\ \approx\ -1.6~\mathrm{rad}\times\left(\frac{L}{1~\mathrm{m}}\right),
\end{equation}
i.e.\ an ${\cal O}(1)$ radian phase shift over meter-scale propagation---well within standard vector-network-analyzer interferometry at cryogenic temperatures. Note that key practical requirements apply: a cryostat compatible with microwave feedthroughs and domain control (ME annealing) so that the sample remains in a single ME domain, consistent with the symmetry assumptions of Sec.\ \ref{subsec:pol_antisym}.

\subsection{Symmetric-traceless representative Cr$_2$O$_3$}

Cr$_2$O$_3$ is the prototypical linear ME. Its diagonal ME coefficient along the trigonal
axis reaches values of order $\alpha_{zz}\simeq 4.31~\mathrm{ps/m}$ near $T\approx 263~\mathrm{K}$, as
reported in experimental studies \cite{Schoenherr2017MeFM}. Since the pure trace sector is propagation-silent, the relevant quantity for wave propagation is the traceless symmetric part, i.e., the anisotropy of the diagonal response. A minimal estimate is obtained by assuming that the in-plane coefficients are of the same order but smaller than $\alpha_{zz}$ (as is typical for uniaxial ME crystals), so that the traceless eigenvalues
$\lambda_a$ entering Sec.\ \ref{subsec:pol_symtr_hk3zero} are of order $\Delta\alpha \sim \alpha_{zz}$ in magnitude.

\begin{table*}[ht]
\caption{Representative estimates for the two propagation-relevant ME sectors.
We quote peak ME coefficients from the literature and compute the phase-speed enhancement for
$\mu_r=1$, $\varepsilon_r=10$ (for scaling, use Eq.~\eqref{eq:dimensionless_strength}).}
\label{tab:estimates}
\begin{ruledtabular}
\begin{tabular}{lccc}
Material (sector) & ME coefficient (SI) & Geometry used & Enhancement \\
\hline
TbPO$_4$ (antisymmetric) &
$\alpha\sim 730~\mathrm{ps/m}$ \cite{Shen2019DyCrO4} &
$\beta=0$ (prop.\ $\perp \alpha_k$) &
$|v|/v_d\approx 1.0024$ \\
Cr$_2$O$_3$ (sym.\ traceless) &
$\alpha_{zz}\simeq 4.31~\mathrm{ps/m}$ \cite{Schoenherr2017MeFM} &
$\hat{k}_3=0$ (uniaxial) &
$|v|/v_{d}\approx 1.0001$  \\
\end{tabular}
\end{ruledtabular}
\end{table*}

In the uniaxial limit Sec.\ \ref{subsec:pol_symtr_hk3zero} (effectively $\lambda_1=\lambda_2$), one has $Q(\phi)=0$ and the
in-plane dispersion reduces to Eq.\ \ref{eq:uni_lim_trless_dis_rel}. For small $|G|\ll 1$, the enhanced branch satisfies
\begin{equation}
r_+\simeq 1+\frac{|G|}{2},
\qquad
|G|\sim \frac{\mu_r v_d\,\Delta\alpha}{2}\simeq \frac{c\,\Delta\alpha}{2\sqrt{\varepsilon_r}}.
\label{eq:Cr2O3_smallG}
\end{equation}
Using $\Delta\alpha\sim \alpha_{zz}\simeq 4.31\times 10^{-12}~\mathrm{s/m}$ and $\varepsilon_r\simeq 10$,
Eq.~\eqref{eq:Cr2O3_smallG} yields
\begin{equation}
|G|\ \sim\ \frac{(3.0\times 10^8)(4.31\times 10^{-12})}{2\sqrt{10}}
\ \sim\ 2.0\times 10^{-4},
\end{equation}
that is,
\begin{equation}
r_+-1\ \sim\ 10^{-4}\quad (100~\mathrm{ppm}).
\end{equation}
This is a small but very realistic target for resonant techniques. For example, at $f=10~\mathrm{GHz}$, Eq.~\eqref{eq:phase_shift} gives a difference in the phase shift of order
\begin{equation}
\Delta\Phi(L)\sim -7\times 10^{-2}~\mathrm{rad}\times\left(\frac{L}{1~\mathrm{m}}\right),
\end{equation}
which could be boosted by longer effective path length (multi-pass interferometry or delay lines), or by working closer to the maximum of $\alpha_{zz}(T)$ (temperature tuning), or most
efficiently, by using a high-$Q$ cavity perturbation measurement where ppm-level frequency shifts are routine.

For convenience, Table~\ref{tab:estimates} summarizes the above estimates. All enhancements scale as $\propto \sqrt{\mu_r/\varepsilon_r}$ through Eq.~\eqref{eq:dimensionless_strength}. Thus, once $\varepsilon_r(\omega)$ and $\mu_r(\omega)$ are specified for a given frequency band, the estimates can be updated immediately.

It should be remarked that the above numbers use quasi-static (or low-frequency) ME coefficients and a nondispersive modeling of
$(\varepsilon,\mu,\alpha)$. For quantitative predictions across wide frequency bands, one must promote $\alpha_{ij}\mapsto \alpha_{ij}(\omega)$ and include losses and Kramers--Kronig constraints. In that case, the group velocity and signal velocity become the relevant causal observables, while the geometric-optics phase speed computed here remains the correct leading-order diagnostic for dispersion engineering and polarization splitting in the low-loss regime. This shall be investigated in a forthcoming paper.

\section{Final remarks}
\label{sec:final}

We provided a propagation-based classification of linear magnetoelectric response in homogeneous dielectrics in the geometric-optics regime. Assuming isotropic dielectric parameters $(\varepsilon,\mu)$ and a generic $3\times 3$ ME matrix $\alpha_{ij}$, we derived the Fresnel-matrix eigenproblem and the quartic dispersion relation for the normalized phase speed $r=v/v_d$. The main result is that the canonical algebraic splitting of $\alpha_{ij}$ into trace, symmetric-traceless, and antisymmetric sectors directly translates into distinct propagation signatures:
(i) the pure-trace sector is propagation-silent at leading order;
(ii) the antisymmetric sector yields a factorized dispersion relation with closed-form branches and, for suitable geometries, explicit phase-speed enhancement relative to $v_d$; and
(iii) the symmetric-traceless sector produces the richest directional dependence and polarization mixing, encoded by a small set of invariants controlling the Fresnel wave surface.

Our estimates indicate that the predicted phase-speed shifts can be accessed with phase-sensitive transmission and resonant techniques in realistic frequency bands, especially when large ME coefficients can be achieved (e.g., via domain control or engineered structures). At the same time, it is important to stress the scope of the present analysis: we assumed a local, linear, nondispersive response. Therefore, a complete causal interpretation of phase-speed enhancement requires incorporating frequency dependence and losses, so that group and front velocities can be analyzed and compared with known causality constraints (cf.\ the broader effective-geometry/NLED literature, including duality-based considerations~\cite{GibbonsRasheed1995NPB}).

Several extensions are natural. First, the case of inhomogeneous and anisotropic dielectrics must be addressed to ensure that the enhancement is still possible when the medium has less symmetry. Some positive results have already been reported in this direction \cite{DeLorenci2025}, but a propagation-based classification is still an open question. Second, dispersive ME response $\alpha_{ij}(\omega)$ should be included, which will make attenuation and polarization rotation measurable and will enable a more direct link to microwave or THz spectroscopy. Second, nonlinear response and higher-order ME coefficients can generate qualitatively new wave-surface features beyond the strictly linear local setting, as indicated in Refs.\ \cite{De_Lorenci_2022,Eduardo2023} for related mechanisms. Finally, in realistic solids, the electromagnetic ME coupling can coexist with elastic degrees of freedom (magnetoelectroelasticity) \cite{SoheLiu2015}, in which case additional constitutive structure may be relevant depending on the experimental protocol.


\acknowledgments
This research is partially supported by \textit{Conselho Nacional de Desenvolvimento Científico e Tecnológico} (grants N.\ 305217/2022-4 and 402093/2022-4) and \textit{Funda\c c\~ao de Amparo \`a Pesquisa do Estado de Minas Gerais} (processes APQ-05207-23 and APQ-02782-25).


\bibliography{ref}

\end{document}